\documentclass[10pt,a4paper]{article}

\usepackage[utf8]{inputenc}
\usepackage{amsmath}
\usepackage{amsfonts}
\usepackage{amssymb}
\usepackage{bbold}
\usepackage[dvips,letterpaper,margin=0.75in,bottom=0.5in]{geometry}
\usepackage[noadjust]{cite}
\usepackage[T1]{fontenc}
\usepackage[utf8]{inputenc}
\usepackage{authblk}
\usepackage{scrextend}
\usepackage{footmisc}
\usepackage{mathrsfs}
\usepackage{bbm} 
\usepackage{float}

\usepackage{tikz}
\usepackage{calc}
\usepackage{caption}
\usepackage{subcaption}
\usetikzlibrary{decorations.markings}
\tikzstyle{vertex}=[circle, draw, inner sep=0pt, minimum size=6pt]
\newcommand{\vertex}{\node[vertex]}

\usepackage{mathtools}
\usepackage{indentfirst}       

\title{Sections and Chapters}
\usepackage{cite}
\newcommand{\ara}{a^{\dagger}}
\newcommand{\ala}{a}
\newcommand{\Pmi}{aa^{\dagger}}
\newcommand{\Pm}[2]{a^{\dagger\/{#1}}a^{#2}}

\newcommand{\Pmb}{a^{\dagger}a}
\newcommand{\Pmn}{a^{\dagger\/n}a^{n}}
\newcommand{\Pmdm}{a^{\dagger\/d-1}a^{d-1}}

\newcommand{\Pmd}{a^{\dagger\/d}a^{d}}

\newcommand{\Rm}[2]{a^{#1}a^{\dagger\/{#2}}}
\newcommand{\Rmm}{a^{m}a^{\dagger\/m}}
\newcommand{\Rmn}{a^{n}a^{\dagger\/n}}
\newcommand{\Rmb}{aa^{\dagger}}
\newcommand{\Rmdm}{a^{d-1}a^{\dagger\/d-1}}

\newcommand{\prn}{P_{n}}
\newcommand{\prnb}{P_{n+1}}
\newcommand{\prm}{P_{m}}
\newcommand{\prmb}{P_{m+1}}

\newcommand{\mpn}{\mathscr{P}_{n}}
\newcommand{\mpm}{\mathscr{P}_{m}}
\newcommand{\mpx}[1]{\mathscr{P}_{#1}}

\newcommand{\ket}[1]{\left| #1 \right>} 
\let\baraccent=\= 
\renewcommand{\=}[1]{\stackrel{#1}{=}} 

\title{Quantum Mechanics on Periodic and Non-Periodic Lattices and Almost Unitary Schwinger Operators}
\author{Metin Arik$^{1,}$\footnote{e-mail:metin.arik@boun.edu.tr}}
\author{Medine Ildes$^{1,}$\footnote{e-mail:medine.ildes@boun.edu.tr}}
\affil{$^{1}$Department of Physics, Bogazici University, Bebek, Istanbul, Turkiye}

\begin{document}

\maketitle

\begin{abstract}
In this work we uncover the mathematical structure of the Schwinger algebra and introduce an almost unitary Schwinger operators which are derived by considering translation operators on a finite lattice. We calculate mathematical relations between these algebras and show that the almost unitary Schwinger operators are equivalent to the Schwinger algebra. We introduce new representations for $M_{N}(C)$ in terms of these algebras. 
\end{abstract}
\section{Introduction}
\indent Quantum mechanics on a finite periodic lattice is a well known subject which has been studied repeatedly since Schwinger's famous 1960 paper \cite{schwinger1960unitary}. He developed the generators of a complete unitary operator basis. Applications of Schwinger approach have been used in quantum optics, quantum communications, quantum probability and Galois quantum systems \cite{durt2005mutually,spengler2012entanglement,durt2010mutually,wootters1989optimal,miquel2002quantum,wootters2004picturing,paz2002discrete,davies1970operational,
vourdas2005galois,vourdas2007quantum,scully1991quantum}. In addition one can find the review of the literature on quantum systems with finite Hilbert space and the link between this theory and the other research fields in Vourdas \cite{vourdas2004quantum}.\\
\indent Schwinger considered a periodic lattice on which the translation operator $U$ is unitary due to the periodicity of the lattice. On such a lattice the position can be again expressed by a unitary operator $V$ such that
\begin{gather}
VU=qUV \hspace{15pt} \text{where} \hspace{15pt} q=e ^{\frac{2\pi i}{d}} \\ \nonumber
V^{d}=U^{d}=1 \hspace{15pt} \text{and} \hspace{15pt} VV^{\dagger}=UU^{\dagger}=1
\end{gather}
\\ where $d$ is the number of points on the periodic lattice. Schwinger chose the integer $d$ to be prime and in this case the relation $q=\exp ^{\frac{2\pi i}{d}}$ can be omitted since it is already implied by the other equations.\\
\indent In \cite{arik2016quantum} it has been shown that a finite lattice has an almost unitary quasi-translation operator $a$ which satisfies
\begin{gather}
\Pmi =1-\Pmdm, \hspace{15pt} \text{and} \hspace{15pt} a^{\dagger d}=0 \\ 
\Pmb =1-\Rmdm, \hspace{15pt} \text{and} \hspace{15pt}  \ala^{d}=0 .\\  \nonumber
\end{gather}
\\ The operators  $\ara$ and $a$ in the above relations can be respectively regarded as the right quasi-translation operator and the left quasi-translation operator since an end point can be translated only in one direction. A point which lies at the right end of the finite lattice can only be translated left end vice versa. Equation (2) gives the minimal set of relations that define the algebra generated by $a$ and $\ara$. The second set of the relations written in equation (3) can be derived using (2). Although the algebra defined by equation (1) and by equation (2) look very different, physically they accomplish basically the same concept. Therefore the exact mathematical relation between them should be unveiled.\\
\indent In this paper we construct the projection operators in terms of the almost unitary translation operators and in terms of the unitary Schwinger operators. Since projection operators play the key role in relations between these two algebras we investigate their properties. Then we are able to write each algebra in terms of the other one. We also find two new representations where the standard basis of  $M_{N}(C)$ is constructed in terms of the projection operators in each algebra. Finally, we formulate an algebra which is related to representing a multi-dimensional lattice in terms of one-dimensional lattices in each direction.
\section{Mathematical structure of the almost unitary translation operators}
\indent In our previous work  \cite{arik2016quantum} projection operators $P_{n}$ were defined as,
\begin{align}
P_{n}=\Pmn, \hspace{15pt} \text{where} \hspace{15pt} P_{0}=1
\end{align}
\\and it was shown that
\begin{align}
\Rmn=1-P_{d-n}.
\end{align}
Thus we can also define another projection operator such as,
\begin{align}
R_{n}=\Rmn, \hspace{15pt} \text{where} \hspace{15pt} R_{0}=1.
\end{align}
Therefore one can easily see the following relations between the projection operators $P_{n}$ and $R_{n}$
\begin{align}
P_{n}=1-R_{d-n} \hspace{15pt} \text{and} \hspace{15pt} R_{n}=1-P_{d-n}.
\end{align}
\\Their properties which are calculated in the appendix are summarized as,
\begin{gather}
P_{n}=\Pmn, \hspace{15pt} P_{0}=1, \hspace{15pt} P_{m}\ara=\ara P_{m-1}, \hspace{15pt} \ala P_{m}=P_{m-1}\ala ,\\ \nonumber
P_{n}P_{m}=P_{j} \hspace{15pt} \text{where} \hspace{15pt} j=\text{max} (n,m), \\  \nonumber
P_{m}a^{n}=a^{\dagger n}P_{m}=0 \hspace{15pt} \text{for} \hspace{15pt} n+m\geq d, \\ \nonumber
\end{gather}
\begin{gather}
R_{n}=\Rmn, \hspace{15pt} R_{0}=1, \hspace{15pt} R_{m}\ara=\ara R_{m+1}, \hspace{15pt} \ala R_{m}=R_{m+1}\ala ,\\ \nonumber
R_{n}R_{m}=R_{j} \hspace{15pt} \text{where} \hspace{15pt} j=\text{max} (n,m), \\ \nonumber
a^{n}R_{m}=R_{m}a^{\dagger n}=0 \hspace{15pt} \text{for} \hspace{15pt} n+m\geq d. \\ \nonumber
\end{gather}
\indent In \cite{arik2016quantum} we have already considered a position space of d points where a particle located at position $ X=\beta n $ is described by the ket vector $ \ket{n}, n=0,1,...,d-1 $ where
\begin{align}
X\ket{n}=\beta n\ket{n} 
\end{align}
\\with $\beta$ as the grid spacing. We can define the position operator as
\begin{align}
X&=\beta \sum\limits_{m=1}^{d-1}P_{m}  ,\\ \nonumber
X&=\beta \{\Pmb+...+\Pmn+...+\Pmdm\}  \nonumber
\end{align}
\\Applying this to $\ket {n}$ one gets the desired result.
\section{The Schwinger algebra in terms of the almost unitary translation operators}
\indent We define unitary operators $U$ and $V$ that cyclically permutes the vectors of a given system in terms of almost unitary operators $\ala$ and $\ara$
\begin{align}
U&=\ara+\ala^{d-1}, \\
V&=\sum\limits_{n=0}^{d-1}q^{n}(P_{n}-P_{n+1}).
\end{align}
\\Then we will show that these definitions satisfy the Schwinger algebra given by equation (1). The first relation we will prove is 
\begin{align}
VV^{k}=\sum_{n=0}^{d-1}q^{(k+1)n}(P_{n}-P_{n+1})
\end{align}
\\ which has the inclusion relation
\begin{align}
V^{d}=VV^{d-1}&=\sum_{n=0}^{d-1}q^{dn}(P_{n}-P_{n+1})  \\ \nonumber
&=\sum_{n=0}^{d-1}(P_{n}-P_{n+1}) \\ \nonumber
&=(1+\Pmb+\Pm{2}{2}+\cdots +\Pmdm)-(\Pmb+\Pm{2}{2}+\cdots +\Pmdm+\Pmd) \\ \nonumber
&=1 
\end{align} 
\\where we used the algebra relation $a^{\dagger d}=0$.
\\We will prove the equation (14) using the method of proof by induction. For $k=1$ we have
\begin{align}
VV&=\sum_{n,m=0}^{d-1}q^{n+m}(\prn-\prnb)(\prm-\prmb) \\ \nonumber
&=\sum_{n,m=0}^{d-1}q^{n+m}(\prn \prm-\prn \prmb-\prnb \prm+\prnb \prmb) \\ \nonumber
&=\sum_{\substack{n,m=0 \\ n\geq m+1}}^{d-1}q^{n+m}(\prn \prm-\prn \prmb-\prnb \prm+\prnb \prmb) \\ \nonumber
&+\sum_{\substack{n,m=0 \\ n=m}}^{d-1}q^{n+m}(\prn \prm-\prn \prmb-\prnb \prm+\prnb \prmb) \\ \nonumber
&+\sum_{\substack{n,m=0 \\ n<m}}^{d-1}q^{n+m}(\prn \prm-\prn \prmb-\prnb \prm+\prnb \prmb). \\ \nonumber
\end{align}
Using the property $\prn \prm=P_{j}$ where $j=$max$(n,m)$, we obtain
\begin{align}
VV&=0+\sum_{\substack{n,m=0 \\ n=m}}^{d-1}q^{n+m}(\prn-\prmb)+0 \\ \nonumber
&=\sum_{n=0}^{d-1}q^{2n}(\prn-\prnb).
\end{align}
\\We assume that for $k=l$ 
\begin{align}
VV^{l}=\sum_{n=0}^{d-1}q^{(l+1)n}(\prn-\prnb).
\end{align}
\\For $k=l+1$, 
\begin{align}
VV^{l+1}&=(VV^{l})V \\ \nonumber
&=\sum_{n=0}^{d-1}q^{(l+1)n}(\prn-\prnb)\sum_{m=0}^{d-1}q^{m}(\prm-\prmb)  \\ \nonumber
&=\sum_{n,m=0}^{d-1}q^{(l+1)n+m}(\prn \prm-\prn \prmb-\prnb \prm+\prnb \prmb) \\ \nonumber
&=\sum_{\substack{n,m=0 \\ n\geq m+1}}^{d-1}q^{(l+1)n+m}(\prn \prm-\prn \prmb-\prnb \prm+\prnb \prmb) \\ \nonumber
&+\sum_{\substack{n,m=0 \\ n=m}}^{d-1}q^{(l+1)n+m}(\prn \prm-\prn \prmb-\prnb \prm+\prnb \prmb) \\ \nonumber
&+\sum_{\substack{n,m=0 \\ n<m}}^{d-1}q^{(l+1)n+m}(\prn \prm-\prn \prmb-\prnb \prm+\prnb \prmb) \\ \nonumber
\end{align}
\\Using the same property $\prn \prm=P_{j}$ where $j=$max$(n,m)$, we obtain
\begin{align}
VV^{l+1}&=0+\sum_{\substack{n,m=0 \\ n=m}}^{d-1}q^{(l+1)n+m}(\prn-\prmb)+0 \\ \nonumber
&=\sum_{n=0}^{d-1}q^{(l+2)n}(\prn-\prnb).\square
\end{align}
\\Then we will show that $VV^{\dagger}=1$ by using definition of $V$
\begin{align}
V=\sum_{n=0}^{d-1}q^{n}(\prn-\prnb). \\ \nonumber
\end{align}
\\Since $\prn=\prn^{\dagger}$ and $q^{\dagger n}=q^{d-n}$ we have
\begin{align}
VV^{\dagger}&=\sum_{m=0}^{d-1}\sum_{n=0}^{d-1}q^{m}q^{d-n}(\prm-\prmb)(\prn-\prnb) \\ \nonumber
&=\sum_{m,n=0}^{d-1}q^{m}q^{d-n}(\prm\prn-\prm\prnb-\prmb\prn+\prmb\prnb) \\ \nonumber
&=\sum_{\substack{m,n=0 \\ m \geq n+1}}^{d-1}q^{m+d-n}(\prm\prn-\prm\prnb-\prmb\prn+\prmb\prnb) \\ \nonumber
&+\sum_{\substack{m,n=0 \\ m=n}}^{d-1}q^{m+d-n}(\prm\prn-\prm\prnb-\prmb\prn+\prmb\prnb) \\ \nonumber
&+\sum_{\substack{m,n=0 \\ m<n}}^{d-1}q^{m+d-n}(\prm\prn-\prm\prnb-\prmb\prn+\prmb\prnb)  \\ \nonumber
&=0+\sum_{m=0}^{d-1}q^{d}(\prm-\prmb)+0 \\ \nonumber
&=(1+\Pmb+\Pm{2}{2}+\cdots+\Pmdm)-(\Pmb+\Pm{2}{2}+\cdots+\Pmdm+\Pmd)  \\ \nonumber
&=1.\square
\end{align}
\\Next, we will show that
\begin{align}
UU^{n}=a^{\dagger n+1}+a^{d-(n+1)} \hspace{15pt} \text{with} \hspace{15pt} n=0,1,\cdots, d-1
\end{align}
\\which implies
\begin{align}
U^{d}&=UU^{d-1} \\ \nonumber
&=a^{\dagger d}+a^{d-d} \\ \nonumber
&=1. \\ \nonumber
\end{align}
\\Our method is proof by induction. For $n=1$ we have
\begin{align}
UU&=(\ara+\ala^{d-1})(\ara+\ala^{d-1}) \\ \nonumber
&=\ara\ara+\ara\ala^{d-1}+\ala^{d-1}\ara+\ala^{d-1}\ala^{d-1}  \\ \nonumber
&=a^{\dagger 2}+(\Pmb)\ala^{d-2}+\ala^{d-2}(\Rmb) \\ \nonumber
&=a^{\dagger 2}+P_{1}\ala^{d-2}+\ala^{d-2}R_{1} \\ \nonumber
&=a^{\dagger 2}+\ala^{d-2}P_{1+d-2}+\ala^{d-2}(1-P_{d-1}) \\ \nonumber
&=a^{\dagger 2}+\ala^{d-2}P_{d-1}+\ala^{d-2}-\ala^{d-2}P_{d-1} \\ \nonumber
&=a^{\dagger 2}+\ala^{d-2}.
\end{align}
\\For $n=l$ we assume that
\begin{align}
UU^{l}=a^{\dagger l+1}+\ala^{d-(l+1)}.
\end{align}
For $n=l+1$ we obtain
\begin{align}
UU^{l+1}&=(UU^{l})U \\ \nonumber
&=(a^{\dagger l+1}+\ala^{d-(l+1)})(\ara+\ala^{d-1}) \\ \nonumber
&=a^{\dagger l+2}+a^{\dagger l+1}a^{d-1}+a^{d-(l+1)}\ara+a^{2d-(l+2)} \\ \nonumber
&=a^{\dagger l+2}+(a^{\dagger l+1}a^{l+1})a^{d-1-(l+1)}+a^{d-(l+2)}(\ala\ara) \\ \nonumber
&=a^{\dagger l+2}+P_{l+1}a^{d-1-(l+1)}+a^{d-(l+2)}R_{1} \hspace{10pt} \text{use (8)} \\ \nonumber
&=a^{\dagger l+2}+a^{d-(l+2)}P_{l+1+d-l-2}+a^{d-(l+2)}(1-P_{d-1}) \\ \nonumber
&=a^{\dagger l+2}+a^{d-(l+2)}P_{d-1}+a^{d-(l+2)}-a^{d-(l+2)}P_{d-1} \\ \nonumber
&=a^{\dagger l+2}+a^{d-(l+2)}. \square
\end{align}
\\The last term in the third line is zero because at most $l=d-2$ according to equation (23). \\
\indent We will obtain $UU^{\dagger}=1$ just by substitution
\begin{align}
UU^{\dagger}&=(\ara+\ala^{d-1})(\ala+a^{\dagger d-1}) \\ \nonumber
&=\Pmb+a^{\dagger d}+a^{d}+\Rmdm  \\ \nonumber
&=P_{1}+R_{d-1}  \\ \nonumber
&=1. \square
\end{align}
\\where we have used (2), (3) and (7).\\
\indent We have the formula for $U$ and $V$, so we will show that $VU=qUV$ just by substitution. Thus left hand side of the formula is equal to
\begin{align}
VU&=\sum_{n=0}^{d-1}q^{n}(\prn-\prnb)(\ara+\ala^{d-1}) \\ \nonumber
&=\sum_{n=0}^{d-1}q^{n}( \prn\ara+ \prn\ala^{d-1}-\prnb\ara-\prnb\ala^{d-1})
\end{align}
\\Since $\prn=\Pmb$ and $P_{0}=1$, $\prn a^{d-1}=0$ except for $n=0$ and $\prnb a^{d-1}=0$ for all $n$. Therefore we obtain
\begin{align}
VU&=\sum_{n=0}^{d-1}q^{n}(\prn \ara-\prnb\ara)+q^{0}a^{d-1} \\ \nonumber
&=P_{0}\ara-P_{1}\ara+\sum_{n=1}^{d-1}q^{n}(\ara P_{n-1}-\ara P_{n})+a^{d-1}  \\ \nonumber
&=\ara-\ara P_{0}+\sum_{n=1}^{d-1}q^{n}(\ara P_{n-1}-\ara P_{n})+a^{d-1} \\ \nonumber
&=\sum_{n=1}^{d-1}q^{n}\ara(P_{n-1}-P_{n})+a^{d-1}
\end{align}
\\where we have used (8). To obtain right hand side of the formula  $VU=qUV$ we calculate,
\begin{align}
UV&=(\ara+\ala^{d-1})\sum_{n=0}^{d-1}q^{n}(\prn-\prnb) \\ \nonumber
&=\sum_{n=0}^{d-1}q^{n}(\ara(\prn-\prnb)+\ala^{d-1}( \prn-\prnb)) \\ \nonumber
&=\sum_{n=0}^{d-1}q^{n}\ara(\prn-\prnb)+\sum_{n=0}^{d-1}q^{n}\ala^{d-1}(\prn-\prnb) \\ \nonumber
&=\sum_{n=0}^{d-1}q^{n}\ara(\prn-\prnb)+\sum_{n=0}^{d-1}q^{n}\ala^{d-1}(1-R_{d-n}-(1-R_{d-n-1}))  \\ \nonumber
&=\sum_{n=0}^{d-1}q^{n}\ara(\prn-\prnb)+\sum_{n=0}^{d-1}q^{n}\ala^{d-1}(R_{d-n-1}-R_{d-n}).
\end{align}
\\ Since $R_{m}=\Rmm$, $a^{d-1}R_{m}=0$ except $m=0$. The element $R_{d-n-1}=R_{0}=1$ for $n=d-1$ so we have,
\begin{align}
UV&=\sum_{n=0}^{d-1}q^{n}\ara( \prn-\prnb)+q^{d-1}\ala^{d-1}  \\ \nonumber
qUV&=\sum_{n=0}^{d-1}q^{n+1}\ara( \prn-\prnb)+q^{d}\ala^{d-1}   \\ \nonumber
&=\sum_{n=1}^{d}q^{n}\ara(P_{n-1}-P_{n})+\ala^{d-1}    \\ \nonumber
&=\sum_{n=1}^{d-1}q^{n}\ara(P_{n-1}-P_{n})+\ala^{d-1} 
\end{align}
\\where we have used the facts that $ q^{d}=1$, $P_{d}=0$ and $\ara P_{d-1}=0$. which is the same result given by equation (30).
\section{Almost unitary operators in terms of the Schwinger algebra}
\indent It is also possible to write the almost unitary operators $a$ and $\ara$ in terms of $U$ and $V$
\begin{align}
\ara=U-(\dfrac{1+V+V^{2}+\cdots+V^{d-1}}{d})U 
\end{align}
\\The expression in the parentheses is called $\mathscr{P}_{0}$. It is shown that $\mathscr{P}_{0}$ is a projection operator (see appendix). We have found that more general projection operators $\mathscr{P}_{n}$ is written as
\begin{align}
\mathscr{P}_{n}=\dfrac{(1+q^{n}V+q^{2n}V^{2}+\cdots+q^{(d-1)n}V^{(d-1)})}{d} 
\end{align}
\\From the definition it is easily seen that
\begin{align}
\mathscr{P}_{d-l}=\mathscr{P}_{-l} \hspace{15pt} \text{and} \hspace{15pt} \mathscr{P}_{d+l}=\mathscr{P}_{l} \hspace{15pt} \text{and} \hspace{15pt} \mathscr{P}_{d}=\mathscr{P}_{0}.
\end{align}
\\The relationships between unitary operators and projection operators are found as
\begin{align}
\mathscr{P}_{n}U^{m}=U^{m} \mathscr{P}_{n+m}, \\ \nonumber
U^{\dagger m}\mathscr{P}_{n}=\mathscr{P}_{n+m}U^{\dagger m} \nonumber
\end{align}
\\ in the appendix. In addition, multiplication of  $\mathscr{P}_{n}$ and  $\mathscr{P}_{m}$ is found in the appendix as
\begin{align}
\mathscr{P}_{n}\mathscr{P}_{m}&=0 \hspace{15pt} \text{for} \hspace{5pt} m\neq n \\ \nonumber
\mathscr{P}_{n}\mathscr{P}_{n}&=\mathscr{P}_{n}  \nonumber
\end{align}
\\In terms of $U$ we may define $\ara$ as 
\begin{align}
\ara \equiv U-\mpx{0}U.
\end{align}
\\We will show that 
\begin{align}
\ara a^{\dagger l}=U^{l+1}(1-\mpx{1}-\mpx{2}-\cdots-\mpx{l+1})
\end{align}
\\which implies the following relation
\begin{align}
a^{\dagger d}&=\ara a^{\dagger d-1} \\ \nonumber
&=U^{d}[1-(\mpx{1}+\mpx{2}+\cdots+\mpx{d})]  \\ \nonumber
&=U^{d}[1-(\mpx{0}+\mpx{1}+\cdots+\mpx{d-1})]  \\ \nonumber
&=\mathbbm{1} (1-1)  \\ \nonumber
&=0
\end{align}
\\where we have used (35) and (A-23). The equation (40) is the second equation of the unitary algebra defined in (2). We will prove the equation (39) by the method of proof by induction. For $n=1$ we have
\begin{align}
\ara\ara &=(U-\mpx{0}U)(U-\mpx{0}U)   \\ \nonumber
&=U(1-\mpx{1})U(1-\mpx{1})  \\ \nonumber
&=U^{2}(1-\mpx{2})(1-\mpx{1})   \\ \nonumber
&=U^{2}(1-\mpx{1}-\mpx{2})
\end{align}
\\where we have used (36) and (37). For $n=l$ we assume that
\begin{align}
\ara a^{\dagger l}=U^{l+1}(1-\mpx{1}-\mpx{2}-\cdots -\mpx{l+1}).
\end{align}
\\Therefore for $n=l+1$ we obtain
\begin{align}
\ara a^{\dagger l+1}&=(\ara a^{\dagger l})\ara \\ \nonumber
&=U^{l+1}(1-\mpx{1}-\mpx{2}-\cdots -\mpx{l+1})U(1-\mpx{1}) \\ \nonumber
&=U^{l+2}(1-\mpx{2}-\mpx{3}-\cdots -\mpx{l+2})(1-\mpx{1}) \\ \nonumber 
&=U^{l+2}(1-\mpx{1}-\mpx{2}+\mpx{2}\mpx{1}-\cdots-\mpx{l+2}+\mpx{l+2}\mpx{1}) \\ \nonumber
&=U^{l+2}(1-\mpx{1}-\mpx{2}-\cdots -\mpx{l+2}).\square
\end{align}
\indent Now we will show that $\Rmb=1-\Pmdm$ in terms of $U$ and $V$, at the left hand side we have,
\begin{align}
\ala \ara &=U^{\dagger}(1-\mpx{0})(1-\mpx{0})U \\ \nonumber
&=U^{\dagger}(1-\mpx{0}-\mpx{0}+\mpx{0}\mpx{0})U \\ \nonumber
&=U^{\dagger}(1-\mpx{0})U  \\ \nonumber
&=U^{\dagger}U(1-\mpx{1})  \\ \nonumber
&=1-\mpx{1} 
\end{align}
\\where we have used (36) and (1). By using (39) we easily obtain
\begin{align}
a^{\dagger d-1}&=U^{d-1}(1-\mpx{1}-\cdots -\mpx{d-1}) \\ \nonumber
&=U^{d-1}(1-(\mpx{1}+\mpx{2}+\cdots +\mpx{d-1}))  \\ \nonumber
&=U^{d-1}(1-(1-\mpx{0})) \\ \nonumber
&=U^{d-1}\mpx{0}
\end{align}
\\where we have used (A-23). Taking hermitian conjugate of this equation and using hermicity property of the projection operators which is shown by (A-21) we get
\begin{align}
a^{d-1}=\mpx{0}U^{\dagger d-1}.
\end{align}
\\Then for the right hand side of  $\Rmb=1-\Pmdm$ we have
\begin{align}
1-\Pmdm &=1-U^{d-1}\mpx{0} \mpx{0} U^{\dagger d-1} \\ \nonumber
&=1-U^{d-1}\mpx{0} U^{\dagger d-1} \\ \nonumber
&=1-U^{d-1}U^{\dagger d-1}\mpx{1} \\ \nonumber
&=1-\mpx{1} \\ \nonumber
&=a\ara
\end{align}
\\at the second line we used the relations given by (37) and at the last line we have used (44). This is the first equation defining the almost unitary algebra given by (2).
\section{New Representations for Basis of $M_{N}(C)$}
\indent The $e_{ij}$ satisfying (46) form the standard basis of $M_{N}(C)$
\begin{align}
e_{ij}e_{kl}=\delta_{jk} e_{il} \hspace{15pt} e_{ij}^{\dagger}=e_{ji}. 
\end{align}
\\In this section we give two new representations of the basis matrices. One of them is written in terms of the almost unitary algebra as
\begin{align}
e_{mn}=a^{\dagger m}R_{d-1}a^{n} \hspace{15pt} \text{where}\hspace{15pt} R_{n}=\Rmn \hspace{15pt}\text{and} \hspace{15pt} m,n=0,1,\cdots, d-1. 
\end{align}
\\ The other one is written in terms of Schwinger $U$ and $V$ operators
\begin{align}
e_{mn}=
\begin{cases}
U^{m-n}\mathscr{P}_{d-n} \hspace{15pt} \text{for} \hspace{15pt} m>n  \\
\mathscr{P}_{d-n}\hspace{15pt} \text{for} \hspace{15pt} m=n  \\
U^{\dagger n-m}\mathscr{P}_{d-n} \hspace{15pt} \text{for} \hspace{15pt} m<n 
\end{cases}
\hspace{5pt} \text{with} \hspace{5pt} \mathscr{P}_{n}=\dfrac{(1+q^{n}V+q^{2n}V^{2}+\cdots+q^{(d-1)n}V^{(d-1)})}{d}.
\end{align}
\\We prove these representations satisfy (48) in the last part of the appendix.
\section{Multi-dimensional lattice in terms of lower dimensional lattices}
\indent Denoting a linear lattice with $d$ elements by $\mathscr{L}_{d}$, we can show the cartesian product $\mathscr{L}_{d_{1}} \bigtimes \mathscr{L}_{d_{2}}$ by the dots in the following figure.
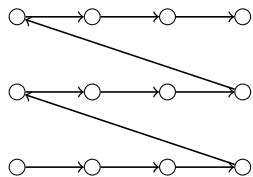
\begin{figure}[H]
\[\begin{tikzpicture}[x=1.0cm, y=1.0cm]
    \vertex (1) at (6,0)  {};
	\vertex (2) at (7,0)  {};
	\vertex (3) at (8,0)  {};
	\vertex (4) at (9,0)  {};
	\vertex (5) at (6,1)  {};
	\vertex (6) at (7,1)  {};
	\vertex (7) at (8,1)  {};
	\vertex (8) at (9,1)  {};
	\vertex (9) at (6,2)  {};
	\vertex (10) at (7,2)  {};
	\vertex (11) at (8,2)  {};
	\vertex (12) at (9,2)  {};
	\path
	
        (1) edge [->,semithick] (2)	    
	    (2) edge [->,semithick] (3)
		(3) edge [->,semithick] (4)
		(5) edge [->,semithick] (6)
		(6) edge [->,semithick] (7)
		(7) edge [->,semithick] (8)
		(9) edge [->,semithick] (10)
		(10) edge [->,semithick] (11)
		(11) edge [->,semithick] (12)
		(4) edge [->,->,semithick](5)
		(8) edge [->,semithick](9)
				
	;	
\end{tikzpicture}\]
\caption*{$4\bigtimes 3$ Lattice}
\end{figure}
Corresponding to this cartesian product of the lattices we have the tensor product of the algebra $\mathscr{A}_{d_{2}} \otimes \mathscr{A}_{d_{1}}$. On the cartesian product shown in the figure the right translation operator corresponds to $\ara\otimes\mathbb{1} $ and the up translation operator corresponds to $\mathbb{1}\otimes\ara$. We denote the (right) translation operator on $\mathscr{A}_{d_{1}}$, $\mathscr{A}_{d_{2}}$, $\mathscr{A}_{d_{1}d_{2}}$ respectively by $\ara _{d_{1}}$, $\ara _{d_{2}}$, $\ara _{d_{2}d_{1}}$ and consider $\mathscr{L}_{d_{1}}$, $\mathscr{L}_{d_{2}}$ as a one one-dimensional lattice as shown by the arrows in the figure. This satisfies an isomorphism
\begin{align}
\mathscr{A}_{d_{2}d_{1}} \xrightarrow{\bigtriangleup} \mathscr{A}_{d_{2}} \otimes \mathscr{A}_{d_{1}}
\end{align}
\\and one can write
\begin{align}
\bigtriangleup (\ara_{d_{2}d_{1}})=\mathbb{1}_{d_{2}}\otimes \ara _{d_{1}}+\ara _{d_{2}}\otimes a^{d_{1}-1}_{d_{1}}.
\end{align}
\\One immediately can check that the action of $\ara _{d_{2}d{1}}$ is given by the arrows in the figure and satisfies the correct algebraic relations.
\\ \indent Similarly, we can express the translation operator for a $d_{1}\times d_{2}$ dimensional periodic lattice by
\begin{align}
\bigtriangleup (U_{d_{2}d_{1}})=\mathbb{1}_{d_{2}}\otimes \ara _{d_{1}}+\ara _{d_{2}}\otimes a^{d_{1}-1}_{d_{1}}+a^{d_{2}-1}_{d_{2}}\otimes a^{d_{1}-1}_{d_{1}}.
\end{align}
\section{Conclusion}
\indent We have shown that the Schwinger algebra can also be given by almost unitary operators which are physically related to the shift operators on a finite lattice. We have named these operators as almost unitary operators because relations 
\begin{align}
UU^{\dagger}=1, \hspace{15pt}  VV^{\dagger}=1 \hspace{15pt} \text{and} \hspace{15pt} VU=qUV \hspace{15pt} \text{where} \hspace{15pt} q=e ^{\frac{2\pi i}{d}} \\ \nonumber
\end{align}
\\are replaced by
\begin{align}
a\ara =1-\Pmdm  \hspace{20pt} \text{and} \hspace{20pt} \ara a=1-\Rmdm
\end{align}
\\ and the terms $\Pmdm$, $\Rmdm$ reflect violation of unitarity for $a$ and $\ara$. For $U$ we have the relation $U^{\dagger}=U^{-1}$ due to the periodic nature of the lattice. Similarly $a$ and $\ara$ can be considered as inverse of each other except at the end points. Note that $a$ and $\ara$ play the role of $U$ and $U^{\dagger}$ where as $V$ can be defined in terms of $a$ and $\ara$. It takes quite an effort to construct $V$ which is given by (13) in terms of $a$ and $\ara$  \\
\indent In usual quantum mechanics where the position and the angular momentum operators have continuous eigenvalues the following equations are equivalent to each other
\begin{gather}
V(p_{0})U(x_{0})=e^{\frac{ix_{0}p_{0}}{\hbar}}U(x_{0})V(p_{0}) \\
[X,P]=i\hbar \\
[X,U(x_{0})]=x_{0}U(x_{0}) \\
\text{here} \hspace{15pt} U(x_{0})=exp(\frac{-ix_{0}P}{\hbar}), \hspace{15pt}  V(p_{0})=exp(\frac{ip_{0}X}{\hbar}) 
\end{gather}
\\For the discrete periodic case $X$ is defined only modulo $2\pi r$. However $U$ and $V$ are well defined. Thus we have only one corresponding equation
\begin{align}
VU=qUV \hspace{15pt} \text{where} \hspace{15pt} q=\exp ^{\frac{2\pi i}{d}}. \\
\end{align}
\\On the other hand for the discrete non-periodic case the position operator $X$, the right quasi-translation operator $\ara$ and the left quasi-translation operator $\ala$ are well defined. Therefore we have only one equation corresponding to (58)
\begin{align}
[X,\ara]= \beta \ara 
\end{align}
\\where $\beta$ is the grid spacing.\\
\indent We have shown how to construct almost unitary translation operators $a$, $\ara$ in terms of $U$, $V$ and vice versa. In addition we have found the relation between basis matrices of $M_{N}(C)$ and the almost unitary operators and the relation between basis matrices of $M_{N}(C)$ and the Schwinger algebra. Furthermore we established an isomorphism between a multi-dimensional and periodic or non periodic linear lattices.  \\
\section*{Acknowledgement}
\indent M.I. thanks Bogazici University for the financial support provided by the Scientific Research Fund (BAP), research project No 11643.
\appendix
\renewcommand{\theequation}{A-\arabic{equation}}      
  \setcounter{equation}{0}  
\section*{Appendix}
\subsection*{Projection operators in terms of  $a$ and $\ara$}
\indent We have found two projection operators $P_{n}$ and $R_{n}$;
\begin{align}
P_{n}=\Pmn \hspace{15pt} \text{and} \hspace{15pt} R_{n}=\Rmn
\end{align}
\\The properties
\begin{align}
P_{m}a^{n}=a^{\dagger n}P_{m}=0 \hspace{15pt} \text{for} \hspace{15pt} n+m\geq d \\ \nonumber
a^{n}R_{m}=R_{m}a^{\dagger n}=0 \hspace{15pt} \text{for} \hspace{15pt} n+m\geq d \\ \nonumber
\end{align}
\\are immediate results of definition of the projection operators and the algebra property $a^{d}=a^{\dagger d}=0$.\\
\indent The following properties have already been proved in \cite{arik2016quantum}.
\begin{align}
P_{n}P_{m}=P_{m} \hspace{15pt} \text{where} \hspace{15pt} m\geq n, \\
P_{m}\ara=\ara P_{m-1}\hspace{15pt} \text{and} \hspace{15pt} P_{m}a=aP_{m+1}.
\end{align}
\\Now, we will prove $R_{n}R_{m}=R_{m}$ where $m\geq n$. Our method is proof by induction.
\\For $n=1$
\begin{align}
R_{1}R_{m}&=(\Rmb)(\Rmm) \\ \nonumber
&=a(\Pmb)\Rm{m-1}{m} \\ \nonumber
&=a(1-\Rmdm)\Rm{m-1}{m}  \\ \nonumber
&=a\Rm{m-1}{m} \\ \nonumber
&=\Rmm \\ \nonumber
R_{1}R_{m}&=R_{m}
\end{align}
\\where we have used $a^{d}=0$. For $n=l$, we assume 
\begin{align}
R_{l}R_{m}=R_{m} \hspace{15pt} \text{for} \hspace{15pt} m\geq l.
\end{align}
\\For $n=l+1$, we have
\begin{align}
R_{l+1}R_{m}&=(\Rm{l+1}{l+1})(\Rmm) \\ \nonumber
&=a(\Rm{l}{l})a^{\dagger}a(\Rm{m-1}{m-1})a^{\dagger} \\ \nonumber
&=aR_{l}(\Pmb)R_{m-1}a^{\dagger} \\ \nonumber
&=aR_{l}(1-\Rm{d-1}{d-1})R_{m-1}a^{\dagger} \\ \nonumber
&=aR_{l}R_{m-1}a^{\dagger}-aR_{l}R_{d-1}R_{m-1}a^{\dagger} \\ \nonumber
&=aR_{l}R_{m-1}a^{\dagger}-aR_{d-1}R_{m-1}a^{\dagger} \\ \nonumber
\end{align}
\\where we have used (2) and (A-6). By using $aR_{d-1}=\Rm{d}{d-1}=0$ and the assumption for $n=l$ with the fact that $l+1\leq m$ implies $l\leq m-1$, so we have $R_{l}R_{m-1}=R_{m-1}$ and $aR_{m-1}\ara =R_{m}$. Therefore
\begin{align}
R_{l+1}R_{m}=R_{m}.\square \\
\end{align}
\\Then we will show that $R_{m}R_{n}=R_{m}$ for $m\geq n$ by using the relation $R_{n}=1-P_{d-n}$ which is equation (7).
\begin{align}
R_{m}R_{n}&=(1-P_{d-m})(1-P_{d-n}) \\ \nonumber
&=1-P_{d-n}-P_{d-m}+P_{d-m}P_{d-n} \\ \nonumber
\end{align}
using $P_{n}P_{m}=P_{m}$ where $m\geq n$ with the fact that $m\geq n$ implies $d-n \geq d-m$
\begin{align}
R_{m}R_{n}&=1-P_{d-n}-P_{d-m}+P_{d-n} \\ \nonumber
&=1-P_{d-m} \\ \nonumber
&=R_{m}.\square
\end{align}
\\By using the last two proofs we conclude that 
\begin{align}
R_{n}R_{m}=R_{j} \hspace{15pt} \text{where} \hspace{15pt} j=\text{max} (n,m).
\end{align}
\\Similarly we will show that $P_{m}P_{n}=P_{m}$ where $m\geq n$ by using $P_{n}=1-R_{d-n}$ which is equation (7).
\begin{align}
P_{m}P_{n}&=(1-R_{d-m})(1-R_{d-n})  \\ \nonumber 
&=1-R_{d-m}-R_{d-n}+R_{d-m}R_{d-n} 
\end{align}
\\using $R_{n}R_{m}=R_{m}$ where $m\geq n$ with the fact that $m\geq n$ implies $d-n \geq d-m$
\begin{align}
P_{m}P_{n}&=1-R_{d-m}-R_{d-n}+R_{d-n} \\ \nonumber
&=1-R_{d-m} \\ \nonumber 
&=P_{m}.\square 
\end{align}
\\This result and equation given by (A-3) can be expressed in one equation as
\begin{align}
P_{n}P_{m}=P_{j} \hspace{15pt} \text{where} \hspace{15pt} j=\text{max} (n,m).
\end{align}
\\Now, we will calculate relations between the projection operator $R_{m}$ and shift operators $a$ and $\ara$.
\begin{align}
R_ {m}\ala&=\Rmm\ala \\ \nonumber
&=\Rm{m}{m-1}(\Pmb) \\ \nonumber
&=\ala(\Rm{m-1}{m-1})P_{1} \\ \nonumber
&=\ala R_{m-1}(1-R_{d-1}) \\ \nonumber
&=\ala R_{m-1}-\ala R_{m-1}R_{d-1}  \\ \nonumber
&=\ala R_{m-1}-\ala R_{d-1}  \\ \nonumber
&=\ala R_{m-1}
\end{align}
\\where we have used (7), (A-11) and the fact that $ \ala R_{d-1}=\Rm{d}{d-1}=0 $
\begin{align}
\ara R_{m}&=\ara \Rmm \\ \nonumber
&=(\Pmb)(a^{m-1}a^{\dagger m-1})\ara \\ \nonumber
&=P_{1}R_{m-1} \ara \\ \nonumber
&=(1-R_{d-1})R_{m-1} \ara \\ \nonumber
&=R_{m-1} \ara-R_{d-1}R_{m-1}\ara \\ \nonumber
&=R_{m-1} \ara-R_{d-1}\ara \\ \nonumber
&=R_{m-1}\ara
\end{align} 
\\where we have used $R_{d-1} \ara=\Pm{d-1}{d}=0$. Let's summarize what we have derived about the projection operators $P_{n}$ and $R_{n}$ up to now;
\begin{align*}
P_{n}=\Pmn, \hspace{15pt} P_{0}=1, \hspace{15pt} P_{m}\ara=\ara P_{m-1}, \hspace{15pt} \ala P_{m}=P_{m-1}\ala , \\ 
P_{n}P_{m}=P_{j} \hspace{15pt} \text{where} \hspace{15pt} j=\text{max} (n,m), \\ \nonumber
P_{m}a^{n}=a^{\dagger n}P_{m}=0 \hspace{15pt} \text{for} \hspace{15pt} n+m\geq d ,\\ \nonumber
\end{align*}
\begin{align*}
R_{n}=\Rmn, \hspace{15pt} R_{0}=1, \hspace{15pt} R_{m}\ara=\ara R_{m+1}, \hspace{15pt} \ala R_{m}=R_{m+1}\ala ,\\ 
R_{n}R_{m}=R_{j} \hspace{15pt} \text{where} \hspace{15pt} j=\text{max} (n,m),  \\ \nonumber
a^{n}R_{m}=R_{m}a^{\dagger n}=0 \hspace{15pt} \text{for} \hspace{15pt} n+m\geq d, \\ \nonumber
\end{align*}
\begin{align*}
P_{n}=1-R_{d-n} \hspace{15pt} \text{and} \hspace{15pt} R_{n}=1-P_{d-n}.
\end{align*}
\subsection*{Projection operators in terms of $U$ and $V$}
\indent We have found two projection operators $\mathscr{P}_{n}$ and $\mathscr{R}_{n}$;
\begin{align}
\mathscr{P}_{n}&=\dfrac{(1+q^{n}V+q^{2n}V^{2}+\cdots+q^{(d-1)n}V^{(d-1)})}{d} \\
\mathscr{R}_{n}&=1-\mathscr{P}_{n}
\end{align}
\begin{align}
(\mathscr{P}_{n})^{2}&=\dfrac{(1+q^{n}V+q^{2n}V^{2}+\cdots+q^{(d-1)n}V^{(d-1)})}{d} \\ \nonumber
& +\dfrac{(q^{n}V+q^{2n}V^{2}+q^{3n}V^{3}\cdots+q^{dn}V^{d})}{d} \\ \nonumber
& +\cdots \\ \nonumber
& + \dfrac{q^{(d-1)n}V^{(d-1)}+q^{dn}V^{d}+q^{(d+1)n}V^{(d+1)}\cdots+2^{2(d-n)}V^{2(d-1)}}{d} \\ \nonumber
&=d\dfrac{(1+q^{n}V+q^{2n}V^{2}+\cdots+q^{(d-1)n}V^{(d-1)})}{d} \\ \nonumber
&=\mathscr{P}_{n}
\end{align}
\\ where we have used $V^{d}=1$ and $q^{d}=1$.\\
\indent By using definition of $\mathscr{P}_{n}$ and Schwinger equation $VU=qUV$, one can easily obtain
\begin{align}
\mathscr{P}_{n}U^{m}=U^{m}\mathscr{P}_{n+m}.
\end{align}
\indent Hermitian conjugate of $\mathscr{P}_{n}$ is easily calculated as,
\begin{align}
(\mathscr{P}_{n})^{\dagger}&=\dfrac{(1+(q^{n})^{\dagger}V^{\dagger}+(q^{2n})^{\dagger}(V^{2})^{\dagger}+\cdots+(q^{(d-1)n})^{\dagger}(V^{(d-1)})^{\dagger})}{d} \\ \nonumber
&=\dfrac{(1+q^{d-n}V^{d-1}+q^{(d-2)n}V^{d-2}+\cdots+q^{n}V)}{d} \\ \nonumber
&=\mathscr{P}_{n}
\end{align}
\\ where we have used the unitary property of $V$ and properties of complex number $q=exp({\frac{2 \pi i}{d}})$. By using hermicity property of $\mathscr{P}_{n}$ we will calculate,
\begin{align}
(\mathscr{P}_{n}U^{m}=U^{m}\mathscr{P}_{n+m})^{\dagger} \\ \nonumber
U^{\dagger m}\mathscr{P}_{n}=\mathscr{P}_{n+m}U^{\dagger m}. \\ \nonumber
\end{align}
\indent The other property of $\mathscr{P}_{n}$ is found by the following steps,
\begin{align}
\mathscr{P}_{0}+\mathscr{P}_{1}+\mathscr{P}_{2}+\cdots+\mathscr{P}_{(d-1)}&=\dfrac{(1+V+V^{2}+\cdots+V^{(d-1)})}{d} \\ \nonumber
& +\dfrac{(1+qV+q^{2}V^{2}+\cdots+q^{(d-1)}V^{(d-1)})}{d} \\ \nonumber
& +\dfrac{(1+q^{2}V+q^{4}V^{2}+\cdots+q^{2(d-1)}V^{(d-1)})}{d} \\ \nonumber
& \cdots \\ \nonumber
& +\dfrac{(1+q^{d-1}V+q^{2(d-1)}V^{2}+\cdots+q^{(d-1)(d-1)}V^{(d-1)})}{d} \\ \nonumber
&=d/d+(1+q+q^{2}+\cdots+q^{(d-1)})V/d \\ \nonumber
& +(1+q^{2}+q^{4}+\cdots+q^{2(d-1)})V^{2}/d \\ \nonumber
& \cdots \\ \nonumber
& +(1+q^{(d-1)}+q^{2(d-1)}+\cdots+q^{(d-1)(d-1)})V^{(d-1)}/d. \\ \nonumber
&=1
\end{align}
\\In the last equation we used the fact that the sum of roots of unity gives zero, 
\begin{align}
(1+q+q^{2})+\cdots+q^{d-1}=0 \hspace{15pt} \text{with} \hspace{15pt} q=exp(\frac{2 \pi i}{d}).
\end{align}
\\Furthermore, when we take unity as $1=exp(2 \pi in)$ where $n$ is a integer, its roots will be $1,q^{n},q^{2n},\cdots,q^{n(d-1)}$. Thus the sum of the terms appear in parentheses in the last lines are equal to $0$.
\\ \indent Then sum of $\mathscr{R}_{n}$ given by equation (A-17) is found easily,
\begin{align}
\mathscr{R}_{0}+\mathscr{R}_{1}+\cdots+\mathscr{R}_{d-1}&=(1-\mathscr{P}_{0})+(1-\mathscr{P}_{1})+\cdots+(1-\mathscr{P}_{d-1}) \\ \nonumber
&=d-1.
\end{align}
\\Now we will show that $\mpn\mpm=0$ unless $n\neq m$ by direct substitution of definition of projection operators
\begin{align}
\mpn\mpm &=(1+q^{n}V+q^{2n}V^{2}+\cdots +q^{(d-1)n}V^{d-1})(1+q^{m}V+q^{2m}V^{2}+\cdots +q^{(d-1)m}V^{d-1})/d^{2} \\ \nonumber
&=\lbrace (1+q^{m}V+q^{2m}V^{2}+\cdots +q^{(d-1)m}V^{d-1}) \\ \nonumber
& +(q^{n}V+q^{n+m}V^{2}+q^{n+2m}V^{3}\cdots +q^{(d-1)m+n}V^{d}) \\ \nonumber
& +(q^{2n}V^{2}+q^{2n+m}V^{3}+q^{2n+2m}V^{4}\cdots +q^{(d-1)m+2n}V^{d+1}) \\ \nonumber
& + \cdots \\ \nonumber
& +(q^{(d-1)n}V^{d-1}+q^{(d-1)n+m}V^{d}+q^{(d-1)n+2m}V^{(d+1)}\cdots +q^{(d-1)(m+n)}V^{2d-2})\rbrace / d^{2}  \\ \nonumber
&=\lbrace 1+(q^{(d-1)m+n}+q^{(d-2)m+2n}+\cdots +q^{(d-1)n+m})V^{d} \\ \nonumber
&  +(q^{m}+q^{n}+q^{(d-1)m+2n}+\cdots +q^{(d-1)n+2m})V \\ \nonumber
& +(q^{2m}+q^{n+m}+q^{2n}+\cdots +q^{(d-1)n+3m})V^{2} \\ \nonumber
& + \cdots \\ \nonumber
& +(q^{(d-1)m}+q^{(d-2)m+n}+\cdots +q^{(d-1)n})V^{d-1}\rbrace / d^{2}
\end{align}
\\with the help of $V^{d+n}=V^{n}$. Now replace $n-m$ by $k$ and use $q^{d}=1$
\begin{align}
\mpn\mpm &=\lbrace (1+q^{k}+q^{2k}+\cdots+q^{(d-1)k}) \\ \nonumber
& + q^{m}(1+q^{k}+q^{2k}+\cdots+q^{(d-1)k})V \\ \nonumber
& + q^{2m}(1+q^{k}+q^{2k}+\cdots+q^{(d-1)k})V^{2} \\ \nonumber
& + \cdots \\ \nonumber
& + q^{(d-1)m}(1+q^{k}+q^{2k}+\cdots+q^{(d-1)k})V^{d-1} \rbrace /d^{2}
\end{align}
\\Furthermore taking unity as $1$, $1+q^{k}+ \cdots +q^{(d-1)k}$ will be equal to equation (A-24) in different order. Hence $(1+q^{k}+\cdots+q^{(d-1)k})=0$. As a result each parentheses in the equation (A-27) are equal to zero except $k=0$ (corresponds $n=m$). At the exception each paratheses sum to $d$. We can summarize our results as
\begin{align}
\mpn\mpm &=0 \hspace{15pt} \text{for} \hspace{5pt} m\neq n\\ \nonumber
\mpn\mpn &=d(1+q^{n}V+\cdots+q^{(d-1)n}V^{d-1})/d^{2}. \\ \nonumber
&=\mpn
\end{align}
\subsection*{$M_{N}(C)$ in terms of $\ala$ and $\ara$}
\indent 
\indent The standard basis of $M_{N}(C)$ is given by the operator $e_{ij}$ which satisfies
\begin{align}
e_{ij}e_{kl}=\delta_{jk} e_{il} \hspace{15pt} e_{ij}^{\dagger}=e_{ji} 
\end{align}
\\$e_{mn}$ can be expressed in terms of $a$ and $\ara$ as,
\begin{align}
e_{mn}=a^{\dagger m}R_{d-1}a^{n} \hspace{15pt} m,n=0,1,\cdots ,d-1 
\end{align}
Then,
\begin{align}
e_{ij}e_{kl}&=(a^{\dagger i}R_{d-1}a^{j})(a^{\dagger k}R_{d-1}a^{l}) \\ \nonumber
&=(a^{\dagger i}a^{j}R_{d-1-j})(R_{d-1-k}a^{\dagger k} a^{l})  \\ \nonumber
&=a^{\dagger i}a^{j}(R_{d-1-j}R_{d-1-k})a^{\dagger k}a^{l} \\ \nonumber
\text{if} \hspace{10pt} j=k \hspace{10pt}&=a^{\dagger i}a^{j}R_{d-1-j}a^{\dagger j}a^{l} \\ \nonumber
&=a^{\dagger i}R_{d-1}a^{l} \\ \nonumber
&=e_{il} \\ \nonumber
\text{if} \hspace{10pt} j>k \hspace{10pt}&=a^{\dagger i}a^{j}R_{d-1-k}a^{\dagger k}a^{l} \\ \nonumber
&=a^{\dagger i}a^{j}a^{\dagger k}R_{d-1}a^{l} \\ \nonumber
&=a^{\dagger i}a^{j}a^{\dagger k}R_{k}R_{d-1}a^{l} \\ \nonumber
&=a^{\dagger i}a^{j-k}R_{d-1}a^{l}  \\ \nonumber
&=0 \\ \nonumber
\text{if} \hspace{10pt} j<k \hspace{10pt}&=a^{\dagger i}a^{j}R_{d-1-j}a^{\dagger k}a^{l} \\ \nonumber
&=a^{\dagger i}R_{d-1}a^{j}a^{\dagger k}a^{l} \\ \nonumber
&=a^{\dagger i}R_{d-1}R_{j}a^{k-j}a^{l}  \\ \nonumber
&=a^{\dagger i}R_{d-1}a^{k-j}a^{l} \\ \nonumber
&=a^{\dagger i}R_{d-1}R_{j}a^{k-j}a^{l} \\ \nonumber
&=0 \nonumber
\end{align}
\\where we have used (9). Thus $e_{ij}e_{kl}=\delta_{jk}e_{il}$. It is the time to check second part of equation (A-29)
\begin{align}
(e_{ij}&=a^{\dagger i}R_{d-1}a^{j})^{*} \\ \nonumber
e_{ij}^{*}&=a^{\dagger j}R_{d-1}^{*}a^{i} \\ \nonumber
e_{ij}^{*}&=a^{\dagger j}R_{d-1}a^{i} \\ \nonumber
e_{ij}^{*}&=e_{ji}
\end{align}
\subsection*{$M_{N}(C)$ in terms of $U$ and $V$}
\indent$e_{mn}$ can be written in terms of operators $U$ and $V$ as
\begin{align}
e_{mn}=
\begin{cases}
U^{m-n}\mathscr{P}_{d-n} \hspace{15pt} \text{for} \hspace{15pt} m>n  \\
\mathscr{P}_{d-n}\hspace{15pt} \text{for} \hspace{15pt} m=n  \\
U^{\dagger n-m}\mathscr{P}_{d-n} \hspace{15pt} \text{for} \hspace{15pt} m<n 
\end{cases}
\hspace{5pt} \text{with} \hspace{5pt} \mathscr{P}_{n}=\dfrac{(1+q^{n}V+q^{2n}V^{2}+\cdots+q^{(d-1)n}V^{(d-1)})}{d}.
\end{align}
\\For $m>n $
\begin{align}
e_{ij}e_{kl}&=U^{i-j}\mpx{d-j}U^{k-l}\mpx{d-l} \\ \nonumber
&=U^{i-j}U^{k-l}\mpx{d-j+k-l}\mpx{d-l} \\ \nonumber
\text{if} \hspace{15pt} j=k \hspace{20pt} &=U^{i-j}U^{j-l}\mpx{d-l}\mpx{d-l} \\ \nonumber
&=U^{i-l}\mpx{d-l} \\ \nonumber
&=e_{il}  \\ \nonumber
\text{if} \hspace{15pt} j\neq k \hspace{20pt} &=U^{i-j}U^{k-l}\mpx{d-j+k-l} \mpx{d-l} \\ \nonumber
&=0
\end{align}
\\Thus $e_{ij}e_{kl}=\delta_{jk}e_{il}$. 
\\Now let us check second part of the equation (A-29) for $m>n$
\begin{align}
(e_{mn}&=U^{m-n}\mpx{d-n})^{\dagger} \\ \nonumber
e_{mn}^{\dagger}&=\mpx{d-n}^{\dagger}U^{\dagger m-n} \\ \nonumber
e_{mn}^{\dagger}&=\mpx{d-n}U^{\dagger m-n}  \\ \nonumber
e_{mn}^{\dagger}&=U^{\dagger m-n} \mpx{d-n-(m-n)}  \\ \nonumber
e_{mn}^{\dagger}&=U^{\dagger m-n} \mpx{d-m} \\ \nonumber
e_{mn}^{\dagger}&=U^{n-m} \mpx{d-m} \\ \nonumber
e_{mn}^{\dagger}&=e_{nm} \\ \nonumber
\end{align}
\\where we have used (A-21), (A-22) and (1). Next for $m=n$ we have
\begin{align}
e_{ii}e_{kk}&=\mpx{d-i}\mpx{d-k} \\ \nonumber
\text{if} \hspace{15pt} i=k \hspace{20pt} &=1  \\ \nonumber
\text{if} \hspace{15pt} i\neq k \hspace{20pt} &=0  \\ \nonumber
\end{align}
\\where we have used (A-28). Thus $e_{ij}e_{kl}=e_{il}$. Now, hermicity of property of the projection operators given by equation (A-21) imply the second part of the matrix algebra given by the equation (A-29).
\\Finally for $m<n$ we have
\begin{align}
e_{ij}e_{kl}&=U^{\dagger j-i}\mpx{d-j}U^{\dagger l-k}\mpx{d-l} \\ \nonumber
&=U^{\dagger j-i}U^{\dagger l-k}\mpx{d-j-l+k}\mpx{d-l} \\ \nonumber
\text{if} \hspace{15pt} j=k \hspace{20pt} &=U^{\dagger j-i}U^{\dagger l-k}\mpx{d-l}\mpx{d-l} \\ \nonumber
&=U^{\dagger l-i}\mpx{d-l} \\ \nonumber
&=e_{il} \\ \nonumber
\text{if} \hspace{15pt} j\neq k \hspace{20pt} &=U^{\dagger j-i}U^{\dagger l-k}\mpx{d-j-l+k}\mpx{d-l} \\ \nonumber
&=0 \\ \nonumber
\end{align}
\\Thus we can conclude that $e_{ij}e_{kl}=\delta_{jk}e_{il}$
\\Now let us check second part of the equation (A-29) for $m<n$ we have
\begin{align}
(e_{mn}&=U^{\dagger n-m}\mpx{d-n})^{\dagger} \\ \nonumber
&=\mpx{d-n}^{\dagger}U^{n-m} \\ \nonumber
&=U^{n-m}\mpx{d-j+j-i} \\ \nonumber
&=U^{n-m}\mpx{d-m} \\ \nonumber
&=e_{nm}.\square \\ \nonumber
\end{align}

\bibliographystyle{unsrt}
\bibliography{bibfile}
\end{document}